\documentclass[aps,prl,amsmath,amssymb,amsfonts,nofootinbib,long]{revtex4}
\usepackage{epsfig,latexsym}

\newcommand{\mq}{m_{\tiny{\rm QCD}}}
\newcommand{\mb}{m_B}
\newcommand{\ov}{\overline}
\newcommand{\bra}{\langle}
\newcommand{\ket}{\rangle}

\newcommand{\mc}{\mathcal}
\newcommand{\gr}{\mathbf}
\renewcommand{\to}{\rightarrow}

\newcommand{\br}{\langle}

\newcommand\G{\mbox{G}}

\newcommand\eV{\mbox{eV}}

\newcommand\GeV{\mbox{GeV}}

\renewcommand\k{\gr{k}}
\newcommand\aem{\alpha_{\rm em}}

\newcommand\Be{\gr{B}_{\rm ext}}

\def\nn{\nonumber \\}
\def\de{\partial}
\newcommand\PP{\mathcal{P}}

\begin{document}

\title{Magnetic Induced Axion Mass}
\author{L. Campanelli$^{1,2,3,4}$}
\email{campanelli@fe.infn.it}
\author{M. Giannotti$^{1,2}$}
\email{giannotti@fe.infn.it}

\affiliation{$^1$INFN - Sezione di Ferrara, I-44100 Ferrara, Italy}
\affiliation{$^2$Dipartimento di Fisica, Universit\`{a} di Ferrara, I-44100 Ferrara, Italy}
\affiliation{$^3$INFN - Sezione di Bari, I-70126 Bari, Italy}
\affiliation{$^4$Dipartimento di Fisica, Universit\`{a} di Bari, I-70126 Bari, Italy}

\date{September 2006}


\begin{abstract}
We study the effect of a uniform magnetic field on the dynamics of
axions. In particular, we show that the Peccei-Quinn symmetry is
explicitly broken by the presence of an external magnetic field.
This breaking is induced by the non-conservation of the magnetic
helicity and generates an electromagnetic contribution to the
axion mass. We compute the magnetic axion mass in one loop
approximation, with no restriction on the intensity of the
magnetic field, and including thermal effects.
\end{abstract}


\maketitle



\section{I. Introduction}

Axions are the pseudoscalar particles predicted by the Peccei and
Quinn (PQ) mechanism~\cite{PQ}, a quite elegant attempt to explain
the smallness of the CP-violation induced by the QCD Lagrangian
(the so-called strong CP problem; for a review see, e.g.,
Ref.~\cite{Kim}). They emerge as Pseudo-Goldstone Boson modes
associated with the global axial PQ-symmetry $U(1)_{\rm PQ}$,
spontaneously broken at the PQ-scale $f_a$. This last is also
known as the PQ- or axion-constant, and characterizes all the
axion phenomenology. More specifically the axion mass is given by
the relation~\cite{Kim}
\begin{equation}
\label{m0} \mq /\eV \simeq 6.2 \times 10^{6} \GeV / f_a,
\end{equation}
(here QCD means that this is induced by interaction with gluons)
while its interactions with fermions are measured by $g_{i}\sim
m_i/f_a$, where $m_i$ represents the fermion mass (e.g.\
$m_e,m_N,...$ for electrons, nucleons, etc.). In addition, though
chargeless, axions interact with photons through the
electromagnetic anomaly:
\begin{equation}
\label{agamma} \mc L_{a\gamma} = \frac14 \, g_{a\gamma} \, a \, F
\widetilde F,
\end{equation}
where $F$ is the electromagnetic field,
$\widetilde F^{\mu\nu} = (1/2) \, \varepsilon^{\mu \nu \alpha
\beta} F_{\alpha \beta}$
its dual, and $g_{a\gamma} = \aem \xi / (2\pi f_a)$. Here, $\aem$
is the electromagnetic fine structure constant and $\xi$ an order
one, model dependent constant. Presently, the PQ-constant's
allowed range is rather narrow~\cite{Raff}: terrestrial
experiments together with astrophysical and cosmological
considerations have in fact excluded all the values of $f_a$ up to
$10^{9} \, \GeV$,
\footnote{In the case of the hadronic axion \cite{KSVZ}, a small
window around $f_a \sim 10^6$GeV can be also permitted (see also
Ref.~\cite{hadronic1}).}
and above $10^{12} \, \GeV$.
\footnote{Attempts to relax the above bounds on the PQ-constant
are discussed in Ref.~\cite{Maurizio}, whereas possibilities of
different cosmological bounds are discussed in Ref.~\cite{Dav86}.}

Because of the axion-photon interaction term (\ref{agamma}), the
phenomenology of axions can be influenced by the presence of an
external electromagnetic field. Probably, the most discussed
result of this well studied interaction is the axion-photon
conversion in an external magnetic field (see, e.g.,
Ref.~\cite{Raff88}), which presently seems the best way to detect
axions by terrestrial experiments~\cite{Sikivie}.

Another interesting consequence of interaction (\ref{agamma}) was
discussed in Ref.~\cite{Mikheev} where it was shown that axions in
a strong magnetic field undergo a finite mass shift. However, in
that computation of the mass shift the temperature effects were
not taken into account. This makes the result in
Ref.~\cite{Mikheev}, though correct, not directly applicable to
the axion phenomenology. In fact, since the axion-photon coupling
is very small, the induced mass can be relevant for the axion
phenomenology only in the case of very intense magnetic fields,
which normally exist only at high temperatures. In this paper, we
reconsider this effect accounting also for the thermal
contribution. As we will show, the result given in
Ref.~\cite{Mikheev} is a very good approximation only for
temperatures well below the electron mass $T \ll m_e$. For higher
temperatures and sufficiently intense magnetic fields the thermal
corrections are relevant.

Interestingly, the resulting mass shift induced by the magnetic
field is independent of the gluon dynamics and, therefore, does
not vanish if the (QCD induced) axion mass (\ref{m0}) is set to
zero. In other words, the external electromagnetic field itself is
responsible for the breaking of the PQ-symmetry, playing a role
similar to the gluon fields. We will discuss this point in Section
2 of this paper. We will show that the magnetic contribution is
indeed due to a topological effect (just as for gluons) which is
expected when the axion dynamics take place in an external
magnetic field. Ultimately, we will associate this effect with the
generation of magnetic helicity, whose important role in
axion cosmology has already been pointed out (see Ref.~\cite{noi2}).
This property makes the magnetic mass shift a true mass term.
Therefore, the terminology ``mass shift'' is perhaps inappropriate.
In the following, we will refer to this contribution as the {\it
magnetic induced axion mass} $\mb$. It contributes to the total
axion mass as
\begin{equation}
m^2 = \mq^2 + \mb^2,
\end{equation}
where $\mq$ is given in (\ref{m0}).

Because of this property, our result applies not only to axions
but to any pseudoscalar particle which interacts with photons as
in Eq.~(\ref{agamma}), independent of its interaction with
gluons. This is the case with arions~\cite{Anselm}, (almost)
massless pseudoscalar particles which do not interact with gluons
at all.
\footnote{Since arions do not interact with gluons, they are not
affected by the color anomaly. Therefore, they do not have a
QCD-induced mass term as in Eq.~(\ref{m0}). The phenomenology of
arions in an external magnetic field has been reconsidered
recently as a tentative to analyze the problem of supernova
dimming in terms of photon-arion oscillations~\cite{Csaki} (see
also Ref.~\cite{Mirizzi}).}
On the other hand, concerning the axion phenomenology, we expect
the contribution $\mb$ to be especially relevant at temperatures
above $\sim 1 \GeV$, where the gluon contribution to the axion
mass is strongly suppressed~\cite{GPY}, or in regions of space and
time where a very strong magnetic field is present. A more
detailed analysis of the phenomenological implications of the
magnetic induced axion mass is in progress.

The paper is organized as follows: In Section~2 we comment on the
origin of the magnetic induced mass term. We then compute the
magnetic induced axion mass in Section~3, discussing the zero
temperature limit as well as the finite temperature result.
Finally, in Section~4 we summarize and give our conclusions.


\section{II. Peccei-Quinn symmetry breaking by an external magnetic field}

In this section we want to discuss some features about the origin of
the axion mass generation due to the external magnetic field. This
discussion is semi-qualitative and based only on symmetry
considerations. The actual computation of the induced
electromagnetic mass term is devoted to the next section. However,
here it will be shown the relevance of the magnetic helicity
for axion dynamics (see also Ref.~\cite{noi,noi2}).

In the PQ-mechanism the axion is defined as the phase of a field
(or, more generally, a linear combination of the phases of
different fields). The Lagrangian is constructed so that it is
invariant under the PQ-symmetry $U(1)_{\rm PQ}$. At pure classical
level the PQ-symmetry is exact and can be read as the invariance
of the effective action $S_{\rm eff}[a(x)]$ [see Eq.~(\ref{3})]
for a constant shift of the axion field $a(x) \to a(x)+c$ ($c=c'
f_a$, with $c'$ a constant).
This symmetry requires the axion interactions to be of the form
$\de_\mu a \, J^\mu$ (where $J^\mu$ is a conserved current), and
prevents the generation of the mass term $m^2 a^2$. If quantum
mechanical effects are taken into account, it results $\de_\mu
J^{\mu}\neq 0$, and the interaction terms with the gluons $(a/f_a)
\, G \widetilde G$ and the electromagnetic field $(a/f_a) \, F
\widetilde F$ emerge. Both $G \widetilde G$ and $F \widetilde F$
are total derivatives. However, because of the non-trivial
topology of the gluon fields, the first one is responsible for the
axion mass term.

Following Shifman, Vainshtein, and Zakharov~\cite{KSVZ} we can
state the problem better defining the quantum effective action for
the axion field, obtained by integrating away all the fields but
the axion:
\begin{equation}
\label{3} e^{-S_{\rm eff}[a(x)]} \: = \int \! D \phi \:
e^{-S[a(x), \phi]} \, ,
\end{equation}
where $\phi$ is a collective name for the gluon, photon and
fermion fields. Since the derivative terms like $\de_\mu a$ are
left invariant for $a \to a+c$, it follows that:
\begin{equation}
e^{-S_{\rm eff}[a(x) + c]} \: = \int \! D \phi \: e^{-S[a(x),
\phi] \; + \; c g \! \int \! d^4x \, G \widetilde G \; + \; c
g_{a\gamma} \! \int \! d^4x \, F \widetilde F}
\end{equation}
($g$ is the coupling with gluons). If we neglect the external
magnetic field we can also neglect the term $\int d^4 x \, F
\widetilde F$ since we know that the topology of the
electromagnetic field is trivial. However, the $\int d^4 x \, G
\widetilde G$ term cannot be neglected and is therefore
responsible for the breaking of the PQ-symmetry, $S_{\rm
eff}[a(x)] \to S_{\rm eff}[a(x)+c]$, and the origin of the axion
mass.
Let us now suppose that an external magnetic field is present in
addition to the radiation one, and reconsider the $\int d^4 x \, F
\widetilde F$ term. We use the relation $F\tilde F = \de^\mu
K_\mu$, where
$K_\mu = \left(\varepsilon_{\mu \alpha \beta \gamma} A^{\alpha}
\de^\beta A^{\gamma}\right) = (K_0,-\gr K)$
and $A^\mu = (A_0,\gr A)$ is the total electromagnetic field,
which includes the external field. Neglecting the contribution of
${\textbf A}$ at the surface of the spatial integration volume
$V$, which implies $\int_{V} \, \! d^3 x \, {\rm div} {\gr K} =
0$, we get
\begin{equation}
\frac{1}{4} \! \int_{t_1}^{t_2} \! d^4 x \, F_{\mu \nu}
\widetilde{F}^{\mu \nu} = H_B(t_2) - H_B(t_1),
\end{equation}
where
\begin{equation}
\label{helicity} H_B(t) = \frac{1}{V} \! \int_{V} \, \! d^3 x \,
{\textbf A} \cdot \nabla \times {\textbf A},
\end{equation}
is known as magnetic helicity or Abelian Chern-Simons term (see,
e.g., Ref.~\cite{Jackiw}). The conclusion is therefore that the
term $\int d^4x \, F \widetilde F$ cannot {\it a priori} be
neglected if the helicity of the system is not conserved.

Interestingly, it was shown a few years ago \cite{Field,noi} that
the dynamics of the axion field (indeed of any pseudo-scalar
field) in an external, uniform magnetic field {\it necessarily}
induces a change of the magnetic helicity. Therefore, the quantity
$H_B(t)$ is not constant and, consequently, $\int d^4x \,F
\widetilde F \neq 0$. The non-vanishing of this term implies the
non-invariance of the classical action for a constant shift of the
axion field,
$S[a(x)+c] - S[a(x)] =
c \, g_{a\gamma} \, \int d^4x \, F \widetilde F \propto \Delta H_B \neq 0$.
This translates in a mass term for the axion field, if also the
quantum effective action is non-invariant $S_{\rm eff}[a(x)+c]
\neq S_{\rm eff}[a(x)]$.
This is the case if all the charged fermions are massive.
On the other hand, if at least one
charged fermion $\psi$ (with electric charge $q_{\psi}$) is
massless, we could redefine the fermion field as
\begin{equation}
\label{chiral} \psi \to \psi' = e^{i\alpha \gamma_5} \, \psi,
\qquad \ov\psi \to \ov{\psi'} = \ov\psi \, e^{i\alpha \gamma_5},
\end{equation}
with $\alpha$ constant. This would introduce a change in the measure
\begin{equation}
d\ov \psi d\psi \to d\ov{\psi'} d \psi' = \: e^{i \alpha'
q_{\psi}^2 \! \int \! d^4x \, F \widetilde F} \: d\ov{\psi} d\psi,
\end{equation}
($\alpha'$ is a constant proportional to $\alpha$) and no changes
in the fermion mass matrix. Therefore, it would always be possible to
choose $\alpha$ so that $(c + \alpha') \int d^4x \, F \widetilde F
= 0$. In other words, if one charged fermion is massless it would
be possible to reabsorb $\int d^4x \, F \widetilde F$ by an
unphysical chiral rotation of that fermion, and then the axion
would not get a mass (this is analogous to what happens to the
standard QCD-induced mass if a quark is massless).

\section{III. Computation of the mass}

In this section we compute the effective, magnetic-induced, axion
mass. Technically, the problem consists in the computation of the
vacuum polarization effects produced by the external (classical)
electromagnetic field. Diagrammatically we consider the
temperature dependent vacuum polarization tensor $\Pi_{\mu\nu}$
with one loop quantum corrections and with an arbitrary number of
interactions with the external field.

For what concerns our conventions we use  natural units with
$\hbar = c = 1$ and the Lorentz-Heaviside convention so that $\aem
= e^2/(4\pi) \simeq 1/137$. The metric tensor is chosen as $g_{\mu
\nu} = {\rm diag}(1,-1,-1,-1)$ and we adopt the often used
notation
\begin{eqnarray}
k_\parallel = (k_0,0,0,k_3), \nn
k_\perp=(0,k_1,k_2,0),
\end{eqnarray}
for the axion momentum $k^\mu = (k_0,\k)$. In addition, we assume
the external magnetic field $\Be$, with intensity $|\Be|=B$, to be
directed along $z$, $\Be = B \hat z$. The total electromagnetic
tensor is
$F_{\rm tot}^{\mu\nu} = F_{\rm ext}^{\mu\nu} + F^{\mu\nu}$,
with $F^{\mu\nu} = \de_\mu A_\nu - \de_\nu A_\mu$ accounting for
the quantum electromagnetic field $A_{\mu}$, and $F_{\rm
ext}^{\mu\nu}$ for the classical external one.

In order to compute the magnetic induced axion mass we need to
consider the exact axion propagator in the external field,
$\Delta(k)$, which is related to the $a \to a$ transition
amplitude in the external field, $M(k)$, by the equation
$\Delta(k )= \Delta_F(k) + \Delta_F(k)M(k) \Delta(k)$,
where $\Delta_F(k)$ is the free axion propagator. The above
equation is shown diagrammatically in Fig.~1. Its formal solution
is
$\Delta^{-1} = k^2 + \mq^2 - M(k)$.
The magnetic axion mass can be calculated in the standard way
taking the limit $\k \to 0$ and solving the equation
$m_B^2 = -{\rm Re} \, M(m_B)$.

We are considering here only the interaction of the axion with the
photon. Of course, also the axion-fermions interactions contribute
to the axion propagator, but the resulting mass shift is small and
disappears in the limit of $\mq \to 0$ \cite{Mikheev}. Therefore,
in the following, it is assumed that $M(k)$ is induced only by
photons (as in Fig.~1).

%
\begin{figure}
\begin{center}
\includegraphics[clip,width=0.8\textwidth]{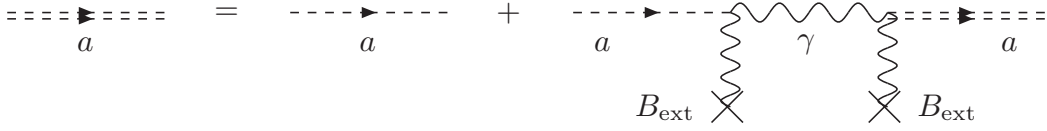}
\caption{Diagrammatic representation of the equation $\Delta(k) =
\Delta_F(k) + \Delta_F(k) M(k) \Delta(k)$ for the exact axion
propagator in the external magnetic field $\Be$.
The sources of this are denoted by crosses.
The dashed lines represent the free axion
propagator, while the double dashed ones represent the exact
propagator. The internal waved line stands for the exact photon
propagator in the external magnetic field.}
\end{center}
\end{figure}
%
%
Formally, the exact photon propagator $\mc D_{\mu\nu}$ is given by
the Dyson relation
\begin{equation}
\label{Dyson}
\mc{D}^{-1}(k) = D^{-1}(k) - \Pi(k),
\end{equation}
where
$D_{\mu\nu} = \bra \mu |D(k)| \nu \ket = k^2 g_{\mu\nu} - \zeta
k_\mu k_\nu$
is the free propagator and $\zeta$ is the gauge parameter.
The polarization tensor $\Pi$ is diagonalizable and its 4
eigenvectors $|e^{(i)}\ket$, $i=1,..,4$ satisfy the orthonormality
$\bra e^{(i)}|e^{(j)} \ket = \delta_{ij}$
 and completeness
$\sum |e^{(i)} \ket \bra e^{(i)}| = 1$
relations \cite{Shabad}. Because of the transversality condition,
$k \cdot \Pi = \Pi \cdot k = 0$, the polarization tensor has a
zero eigenmode directed along $k$. To keep the same notation as in
Ref.~\cite{Shabad}, we label this as the fourth eigenvector
$e_{\mu}^{(4)} = \bra \mu|e^{(4)} \ket = k_\mu / |k|$. So we find
\begin{equation}
\label{Pi} \Pi = \sum_{i=1}^3 \lambda_i |e^{(i)} \ket \br
e^{(i)}|,
\end{equation}
where we have indicated with
\begin{equation}
\label{lambda}
\lambda_i = \br e^{(i)}|\Pi|e^{(i)} \ket = e_{\mu}^{*(i)}
\Pi_{\mu\nu}e_{\nu}^{(j)}
\end{equation}
the eigenvalue associated with $|e^{(i)}\ket$. Of course, it results
$\lambda_4=0$. Using the orthonormality and completeness
relations, it is easy to invert Eq.~(\ref{Dyson}) to get a formal
expression for the exact photon propagator
\begin{equation}
\label{prop} \mc D = \sum_{i=1}^3 \frac{|e^{(i)} \ket \br
e^{(i)}|}{k^2 - \lambda_i} + (1 - \zeta) \frac{|e^{(4)} \ket \br
e^{(4)}|}{k^2} \: .
\end{equation}
In an external magnetic field the relevant term of the
axion-photon interaction (\ref{agamma}) can be conveniently
written as
\begin{equation}
\label{agamma2} \mc L_{a\gamma} = g_{a\gamma} a \,
\widetilde{F}_{\rm ext}^{\mu\nu} \, \de_{\mu} A_{\nu}.
\end{equation}
Observing that the vector
$e^{(3)}_{\mu} = k^{\nu} \widetilde F^{\rm ext}_{\nu\mu}/(|k_\parallel| B)$
directed along $\Be$ is an eigenvector because of the azimuthal
symmetry induced by $\Be$, this interaction takes (in momentum
space) the particularly simple form

\begin{equation}
\label{lagrangiank} \mc L_{a\gamma} = -i g_{a\gamma} \,
|k_\parallel| B \, a A^\mu e_{\mu}^{(3)}.
\end{equation}
Using the orthogonality between the eigenvectors together with
Eq.~(\ref{prop}), we find
\begin{equation}
M(k) = -\frac{g_{a\gamma}^2 B^2 k_{\parallel}^2}{k^2 -
\lambda_{3}} \, ,
\end{equation}
where the eigenvalue $\lambda_3$ is given by Eq.~(\ref{lambda}).
In the limit $\k \to 0$, it results $\lambda_3 = \Pi_{33}$. Let us
define the function $\PP(k_0)$ such that
\begin{equation}
\frac{\aem}{3\pi} \, \PP(k_0) \equiv -\frac{\Pi_{33}(k_0, \k \to
0)}{k_0^2} \, ,
\end{equation}
where we have introduced the factor $\aem/3\pi$ for convenience.
It results
\begin{equation}
M(k_0) = -\frac{g_{a\gamma}^2 B^2}{1 + (\aem/3\pi) \PP(k_0)} \, .
\end{equation}
The one loop polarization tensor at finite temperature in an
arbitrary magnetic field was computed using the imaginary time
formalism in Ref.~\cite{Alexandre}. In the imaginary time
formalism, the (Euclidean) photon energy is quantized and
expressed in terms of the so-called Matsubara frequencies:
$k_4 = -i k_0 \to \omega_n = 2 \pi n T$,
with $n$ integer.
For simplicity, we consider only the contribution of the electrons
in the fermion loop since this is the leading one as electrons are
the lightest charged fermions. In any case, the contribution from
the other fermions can be easily included.  The electron energy in
the fermion loop is quantized as well,
$p_4 = -i p_0 \to \omega_l = (2l+1) \pi T$,
with $l$ integer. The function $\PP(\omega_n)$ is given in
Ref.~\cite{Alexandre}. However, for our purpose, the computation
of the effective axion mass can be simplified observing that in
all the phenomenologically relevant cases the axion mass is always
smaller than the fermion masses $m_B \ll m_e$. This assumption
will be verified {\it a posteriori}. Therefore, we can safely
consider the limiting value
$\PP\equiv\PP(\omega_n/m_e \rightarrow 0)$ in place of
$\PP(\omega_n)$. So finally for the magnetic-induced axion mass we
find
\begin{equation}
\label{dm} \mb^2 \simeq \frac{g_{a\gamma}^2 B^2}{1 + (\aem/3\pi)
\PP} \, .
\end{equation}
From Ref.~\cite{Alexandre} we have
\begin{equation}
\label{P} \PP = \int_0^\infty \frac{du}{u} \: e^{-u m_e^2}
                \, \left\{ 3\sqrt{\pi} \sqrt{u} \, T \!\!
                \sum_{l = -\infty}^{\infty} \!
                \left[e^{-u \omega_l^2} \! \left(\! 1- \frac{2}{3} \,
                \omega_l^2 u \! \right) \! \right] \!
                \ov u\coth \ov u -1 \! \right\},
\end{equation}
where $\ov u = eBu$. Observe that the above expression is
convergent and the last term in the right hand side of
Eq.~(\ref{P}) represents the contact term which removes the
ultraviolet divergence ($u \to 0$). The normalization chosen gives
$\mb = 0$ in the zero temperature and zero magnetic field limit.

\subsection{IIIa. Low Temperature Limit}

Before a complete analysis of expression (\ref{P}) let us consider
its low temperature limit $\PP_0$. Actually, the computation of
$\PP_0$ is straightforward since in the limit $T\to 0$ the sum
over the Matsubara frequencies of the electrons can be substituted
by an integral over the (Euclidean) electron energy: $T \sum_l \to
(2\pi)^{-1} \! \int d p_4$.
The result,
\begin{equation}
\label{mikheev1} \PP_0 = \lim_{T \to 0} \PP = \int_0^\infty
\frac{du}{u} \, e^{-u m_e^2} \left[\ov u \coth \ov u -1 \right],
\end{equation}
is in agreement with the existent literature (see
Ref.~\cite{Mikheev}).
To have a better comparison with the non-zero temperature result
it is convenient to introduce the dimensionless variable
\begin{eqnarray}
\label{new} \lambda = \frac{B}{B_c} = \frac{eB}{m_e^2} \, ,
\end{eqnarray}
so that
\begin{equation}
\label{Ps} \PP_0 = \int_0^\infty \! ds e^{-s/\lambda}
\left[\coth{s} - \frac{1}{s} \right].
\end{equation}
(The parameter $B_c = m_e^2/e \simeq 4.4 \times 10^{13} \mbox{G}$
is known as the critical or the ``Schwinger value'' for the
magnetic field.) A good approximation of the above integral (see
Fig.~2) is given by the following expression:
%
%
\begin{figure}
\noindent\includegraphics[width=8cm,angle=0]{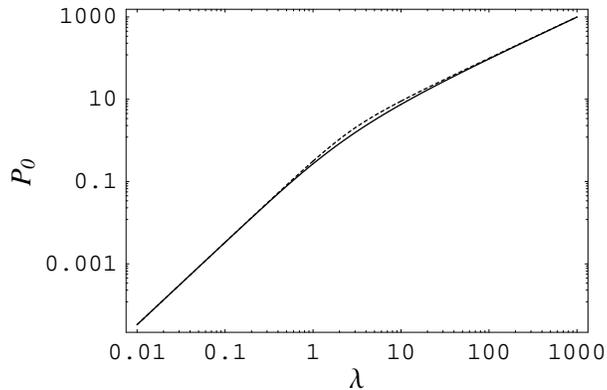}
\caption{\protect\small The continuous line represents the
function $P_0$ in Eq.~(\ref{Ps}), while the dotted line is its
approximate profile Eq.~(\ref{approxP0}).}
\end{figure}
%
%
\begin{equation}
\label{approxP0}
\PP_0 \simeq \lambda \coth{\lambda} - 1.
\end{equation}
In particular, the limit $\lambda\ll 1$ ($B\ll B_c$) gives $\PP_0
\simeq \lambda^2/2$, whereas in the opposite limit $\lambda \gg 1$
($B\gg B_c$) we find $\PP_0 \simeq \lambda$. Taking into account
Eq.~(\ref{dm}), this last limit gives
\begin{equation}
\label{deltamBbis} \mb^2 \simeq \frac{g_{a\gamma}^2 B^2}{1 +
(\aem/3\pi) (B/B_c)} \, ,
\end{equation}
which for $B/B_c \gg 3\pi/\aem \sim 10^3$ gives in turn the result
\begin{equation}
\label{deltamB} \mb \, \simeq \: \frac{3\pi}{\aem} \:
g_{a\gamma}^2 B_c B \: \simeq \: 5.8 \, \xi \left(
\frac{B}{10^{23} \mbox{G}} \right )^{\! 1/2} \frac{10^6 \GeV}{f_a}
\: \eV
\end{equation}
presented in Ref.~\cite{Mikheev} and used in Ref.~\cite{noi3}.

\subsection{IIIb. Finite Temperature Results}

In order to include the thermal effects it is necessary to
calculate the sum over the Matsubara frequencies in Eq.~(\ref{P}):
\begin{equation}
\sum_{l=-\infty}^{\infty} \!\! e^{-u \omega_l^2} \!\left(\! 1 -
\frac{2}{3} \; \omega_l^2 u \! \right)\! .
\end{equation}
This sum can be conveniently expressed in terms of the the Jacobi
theta function of second kind \cite{Abramowitz},
\begin{equation}
\vartheta_2(w,q) = 2q^{1/4} \sum_{n=0}^{\infty} q^{n(n+1)} \cos
[(2n+1)w].
\end{equation}
Indeed, observing that
\begin{eqnarray}
&& \sum_{l = -\infty}^{\infty} e^{-u \omega_l^2} = \vartheta_2
(0,e^{-4\pi^2 u T^2}),
\\
&& \sum_{l = -\infty}^{\infty} \omega_l^2 e^{-u \omega_l^2} = -
\frac{d}{du} \, \sum_{l = -\infty}^{\infty} e^{-u \omega_l^2},
\end{eqnarray}
we have
\begin{equation}
\label{sum} \sum_{l=-\infty}^{\infty} \!\! e^{-u \omega_l^2}
\!\left(\! 1 - \frac{2}{3} \; \omega_l^2 u \! \right) = \left( 1 +
\frac{2}{3} \, u \, \frac{d}{du} \right) \! \vartheta_2
(0,e^{-4\pi^2 u T^2}).
\end{equation}
To proceed further, we define the function
\begin{equation}
\label{F} \mathcal{F}(x) = \sqrt{\frac{x}{\pi}} \left( 3 + 2x
\frac{d}{dx} \right) \! \vartheta_2(0,e^{-4x}),
\end{equation}
whose graph is shown in Fig.~3.
%
%
\begin{figure}
\noindent\includegraphics[width=8cm,angle=0]{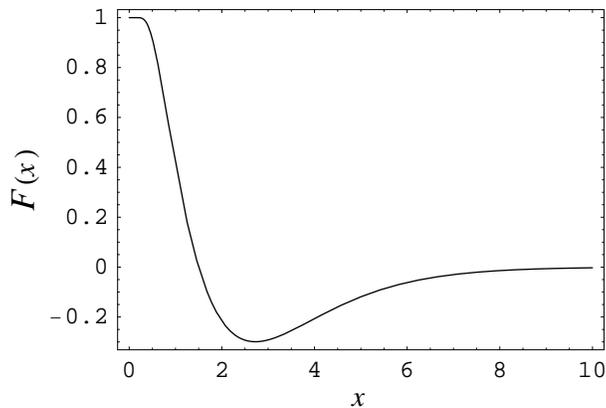}
\caption{\protect\small The function $\mc F(x)$ defined in
Eq.~(\ref{F}).}
\end{figure}
%
%
Taking into account Eqs.~(\ref{sum}) and (\ref{F}), we re-cast
Eq.~(\ref{P}) in the following form:
\begin{equation}
\label{Pbars} \PP = \int_0^\infty \! ds \, e^{-s/\lambda}
\left[\mc F(st^2) \coth{s} - \frac{1}{s} \right],
\end{equation}
where we have defined
\begin{equation}
t = \frac{\pi T}{\sqrt{eB}} \, .
\end{equation}
Equation~(\ref{Pbars}) is very similar to Eq.~(\ref{Ps}), except
for the presence of the function $\mc F(st^2)$. It is important to
notice that the temperature enters only through $\mc F(st^2)$, and
only as $st^2$. Let us comment this result. First of all we see
from Fig.~3 that $\mc F(x) \sim 1$ for $x$ sufficiently smaller
than $1$. In addition, in Eq.~(\ref{Pbars}) the parameter
$\lambda$ has the role of a cutoff. Therefore, in the relevant
integration region, $0 < s < few \times \lambda$, it results $st^2
< few \times \lambda t^2 = (\pi T/m_e)^2$. Then, if $T \ll m_e$,
we can set $\mc F \sim 1$ and consequently we find the result
(\ref{Ps}). On the other hand, for $T \gg m_e$ and $B < T^2$, we
find the asymptotic expansion of $\mathcal{P}$:
\begin{equation}
\label{cosmo5} \mathcal{P} \simeq \gamma + \ln c - \ln \!
\left(\frac{\pi T}{m_e} \right)^{\!2} \! ,
\end{equation}
where $\gamma \simeq 0.577$ is the Euler's constant, $c \simeq
2/3$ is a numerical constant. If $B > T^2$, the above equation is
still valid with $c$ a slowly decreasing function of $B/T^2$ (see
Fig.~4).
\\
Our numerical results are shown in Fig.~5 which represents $\PP$
given in Eq.~(\ref{Pbars}) as a function of the normalized
temperature $T/m_e$, for different values of the parameter
$\lambda = B/B_c$. Also represented is the asymptotic expansion
Eq.~(\ref{cosmo5}). A look at Fig.~5 shows that for a large range
of values for the external magnetic field and temperatures, the
term $(\aem/3\pi) \PP$ is much smaller than 1. In particular, this
is the case for high temperatures ($T \gg m_e$) whatever the
intensity of the magnetic field is, or for low temperatures ($T
\ll m_e$) and magnetic fields below the critical value ($B <
B_c$). In these cases, from Eq.~(\ref{dm}) we get that the axion
mass is well approximated by
\begin{equation}
\label{mBT} \mb \simeq g_{a\gamma} B.
\end{equation}
In order to quantify the thermal effects on the axion mass, we
compare the above result with the axion mass as calculated in the
zero-temperature limit and for large magnetic field (see Section
IIIa). From Eqs.~(\ref{deltamB}) and (\ref{mBT}) we get
\begin{equation}
\frac{\mb^2(T,B)}{\mb^2(0,B)} \simeq \frac{\aem}{3\pi} \,
\frac{B}{B_c} \, .
\end{equation}
%
%
\begin{figure}
\noindent\includegraphics[width=8.0cm,angle=0]{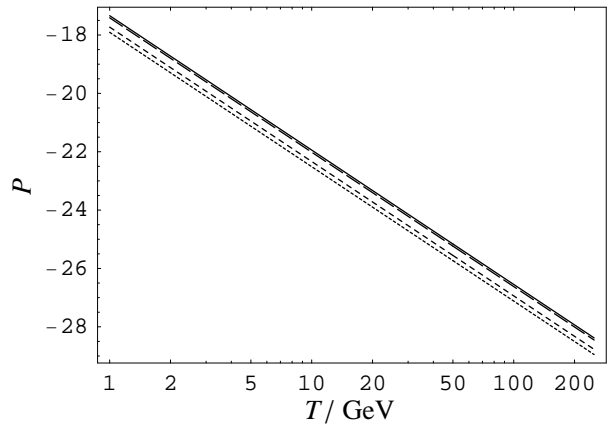}
\caption{The function $\mathcal{P}$ defined in Eq.~(\ref{Pbars})
versus the temperature for different values of the ``normalized''
magnetic field $b = B/T^2$. Dotted line: $b=100$; Dashed line:
$b=50$; Long-dashed line: $b=10$; Solid line: $b=1$. For $b
\lesssim 5$, all curves collapse into the line $b=1$, which is
well approximated by Eq.~(\ref{cosmo5}).}
\end{figure}
%
%
\begin{figure}
\includegraphics[width=8cm,angle=0]{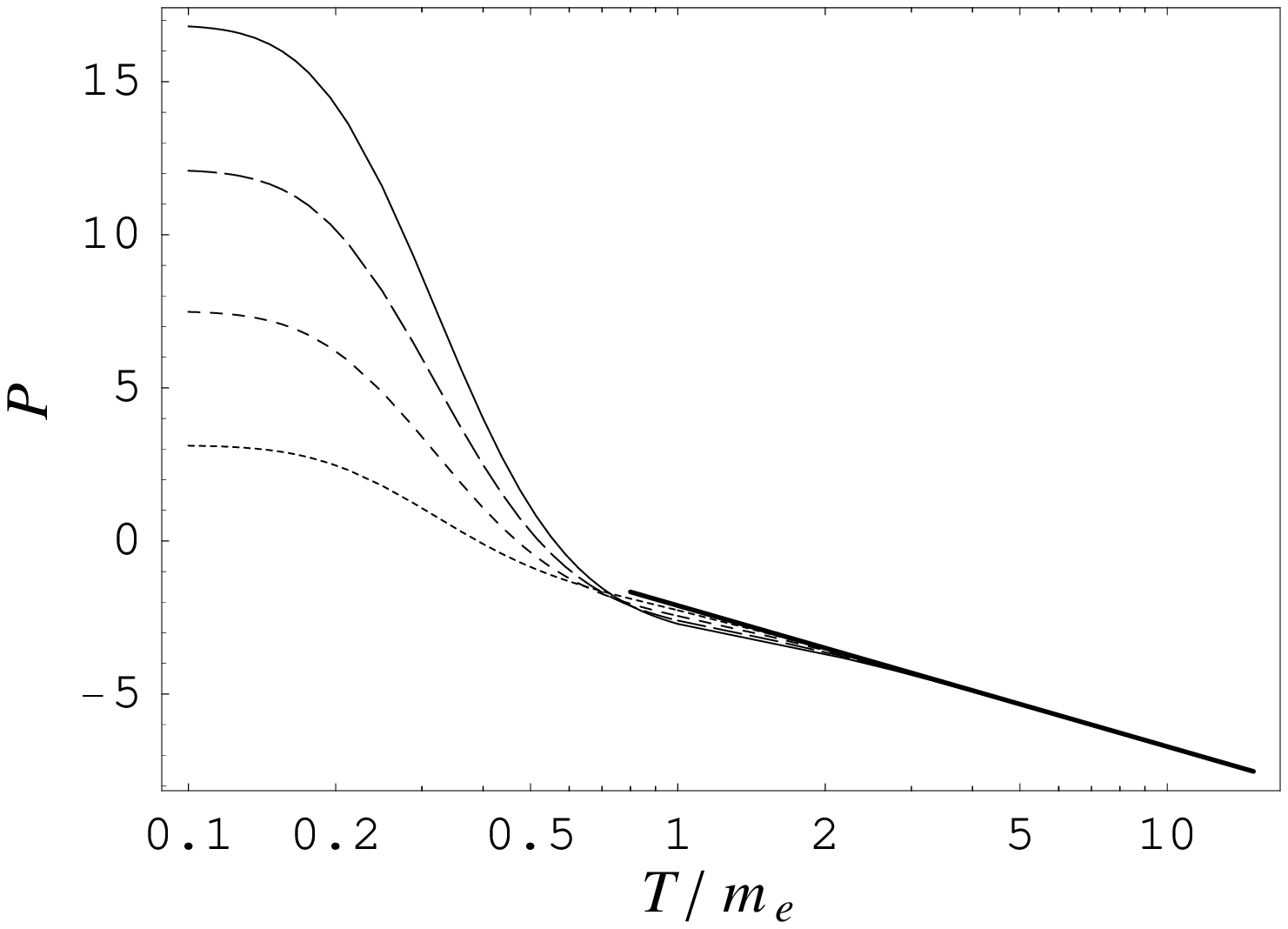}
\includegraphics[width=8cm,angle=0]{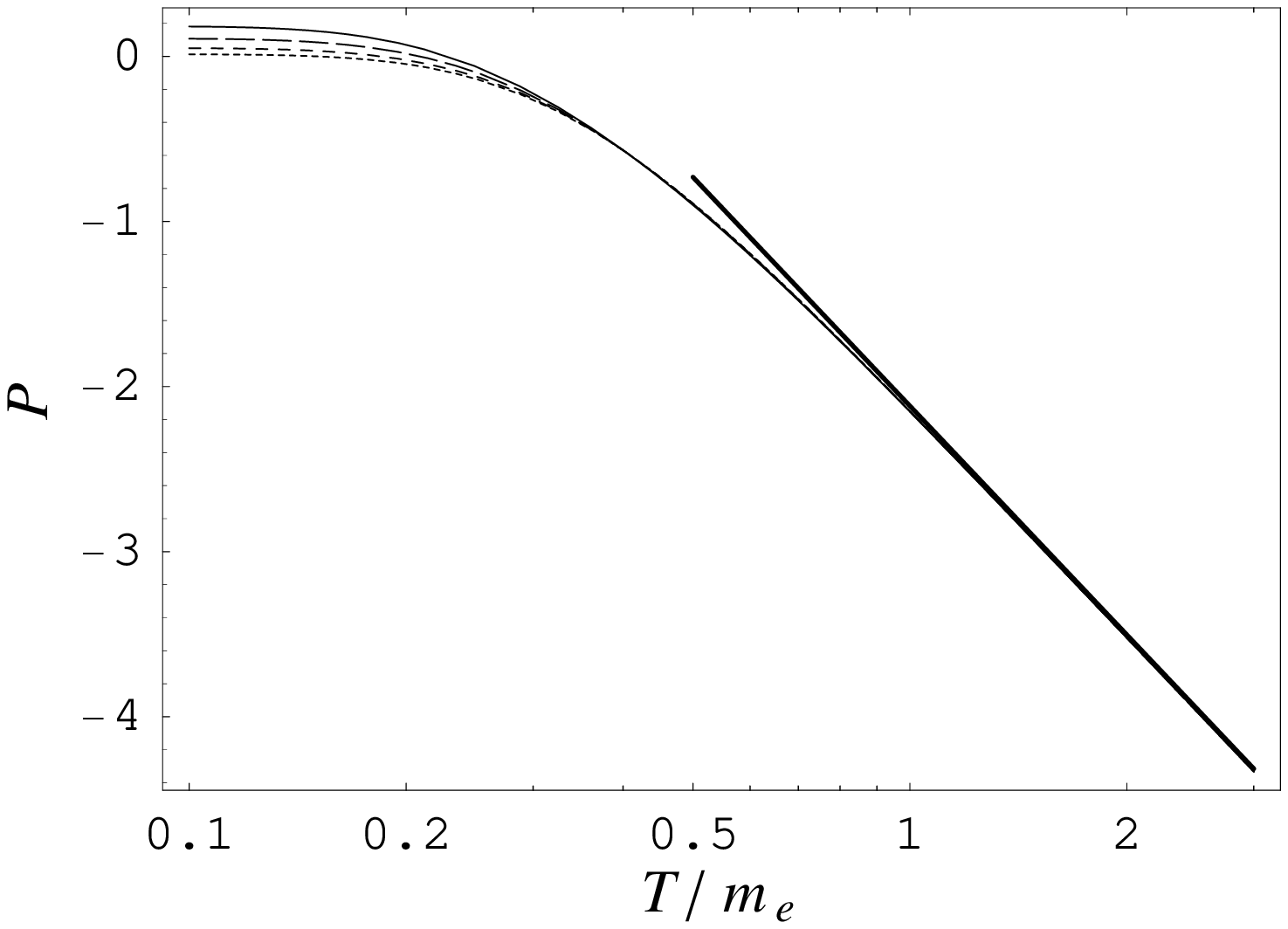}
\caption{\protect\small Numerical representation of the function
$\PP$ given in Eq.~(\ref{Pbars}) as a function of the normalized
temperature $T/m_e$ for different values of the parameter $\lambda
= B/B_c$. Left panel. From the top to the bottom: $\lambda =
20,15,10,1$. Right panel. From the top to the bottom: $\lambda =
0.8,0.6,0.4,0.2$. The tick solid line in both panels represents
the asymptotic expansion (\ref{cosmo5}).}
\end{figure}
%
%
Therefore, for $T \gg m_e$ and $B \gg B_c$, we find that the axion
mass is strongly influenced by temperature effects. In particular,
thermal effects give rise to in increment of the axion mass with
respect to the zero temperature limit.

Before closing this section, a comment is in order. The complete
result in Eq.~(\ref{cosmo5}) is somewhat puzzling, since it gives
a singular limit when $m_e \to 0$, whereas we would expect $\mb
\to 0$ in the chiral limit, as discussed in Section II. However,
it is easy to understand that this limit can not be considered in
our discussion. In fact, the approach that we have considered is
based on the perturbative expansion of Fig.~1, and loses clearly
its validity if $M(k)$ is arbitrarily large, as in the case of
very small electron mass. Indeed, our result is based on an even
stronger assumption, which is that the electron mass is always
much larger than the axion mass. Therefore, the result (\ref{dm})
is valid only if the inequality
\begin{equation}
\left(\frac{g_{a\gamma} B}{m_e}\right)^{\! 2} \, \frac{1}{1 +
(\aem/3\pi) \mathcal{P}} \ll 1
\end{equation}
is verified, which is not the case in the chiral limit.
However, for the standard value of the electron mass, the above
inequality is essentially always verified in any physical
interesting case.
In fact, even in the ``worst'' case $\mathcal{P} \ll 1$ (which
refers to the case $T \ll m_e$ and $B \ll B_c$, or to the case $T
\gg m_e$), the above condition translates in
\begin{equation}
B \ll 10^{28} \, \frac{f_a}{10^9 \GeV} \: \G,
\end{equation}
which is satisfied, even in the most conservative case $f_a = 10^9
\GeV$, by any cosmological and astrophysical magnetic field
present in the universe below the electroweak phase
transition~\cite{Magnetic}.


\section{IV. Conclusions}

The phenomenology of the axion in an external magnetic field has
been deeply studied in the last few decades. Nowadays the
relevance of an in depth understanding of the effects of axion
interaction in an external magnetic field to astrophysics and
cosmology is clear. In this paper we have further investigated
this problem computing the axion mass induced by an external
magnetic field. The origin of this phenomenon is related to the
generation of magnetic helicity,
induced by the axion itself. We computed the
resulting induced mass in one loop approximation, accounting for
the thermal corrections, and with no restriction on the intensity
of the magnetic field. Our result is given in Eqs.~(\ref{dm}) and
(\ref{Pbars}) (see also Figs.~4 and 5), and indicates that thermal
effects are not relevant for temperatures significantly below the
electron mass. For $T \ll m_e$ our result is indeed in agreement
with the existent literature. However, for higher temperatures,
the axion mass is strongly influenced by temperature effects if
the external magnetic field is much stronger than the Schwinger
value, $B \gg B_c \sim 10^{13} \G$. In this case, thermal effects
give rise to an increment of the axion mass with respect to the
zero temperature limit, which is proportional to the magnetic
field: $\mb^2(T,B)/\mb^2(0,B) \simeq (\aem/3\pi) \, (B/B_c)$.

It is clear that the value of the magnetic field that could
induce a sizable change of the axion dispersion relation is
very high, and certainly well above the possibility of the present
terrestrial experiments. However, in an astrophysical or
cosmological contest the magnetic induced mass could play an
important role. This topic will be the object of future
investigations.

\vspace*{0.5cm}

\begin{acknowledgments}
We would like to thank P. Cea, V. Laporta, R. D. Peccei, G. G.
Raffelt, and M. Ruggieri for helpful discussions.
\end{acknowledgments}


\end{document}